\documentclass[aps,twocolumn,showpacs,amsmath,amssymb,pra,superscriptaddress,floatfix,longbibliography]{revtex4-2}
\usepackage{lipsum}
\usepackage{hyperref}
\usepackage{bm}
\usepackage{physics}
\usepackage{braket}
\usepackage{amsmath}
\usepackage{amssymb}
\usepackage{mathtools}
\usepackage{graphicx}
\usepackage{dcolumn}
\usepackage{bm}
\usepackage{multirow}
\usepackage{comment}
\usepackage{leftidx}
\usepackage{subcaption}
\usepackage{geometry}
\usepackage{svg}

\usepackage{epstopdf}
\usepackage[normalem]{ulem}
\usepackage{color}
\usepackage{xcolor}
\usepackage{float}
\usepackage{capt-of}
\usepackage{mwe}
\usepackage{amsfonts}
\usepackage[capitalise]{cleveref}
\usepackage[shortlabels]{enumitem}
\usepackage[german,english]{babel}
\usepackage{gensymb}
\usepackage[thinc]{esdiff}
\usepackage{url}
\graphicspath{ {./Images/Free}, {./Images/bound} }
\setlist[enumerate,1]{label={(\roman*)}}
\begin{document}

\title{X-ray diffraction from chiral molecules with twisted beams}

\author{Akilesh Venkatesh}
\email{akilesh.venkatesh1[at]gmail[dot]com}
\affiliation{Chemical Sciences and Engineering Division, Argonne National Laboratory, Lemont, Illinois 60439, USA}

\author{Phay J. Ho}
\affiliation{Chemical Sciences and Engineering Division, Argonne National Laboratory, Lemont, Illinois 60439, USA}

\author{J{\'e}r{\'e}my R. Rouxel}
\email{jrouxel@anl.gov}
\affiliation{Chemical Sciences and Engineering Division, Argonne National Laboratory, Lemont, Illinois 60439, USA}

\date{\today}

\begin{abstract}
Structured x-rays carrying an orbital angular momentum break spatial inversion symmetry and have been proposed as probes of chirality. 
We theoretically investigate twisted non-resonant x-ray diffraction from chiral molecules and demonstrate that no dichroic signal can arise from randomly oriented molecules, irrespective of the beam spatial profile. 
However, a dichroic response emerges for oriented molecules. 
Our results establish the beam and sample conditions for which a measurable dichroic scattering signal survives axial and focal averaging.
\end{abstract}

\maketitle

\section{Introduction}

Chirality, the non-superimposability of an object on its mirror image, is a structural property that plays an important role in biochemistry \cite{gal2013molecular}, stereochemistry \cite{gal2013molecular} and asymmetric catalysis \cite{walsh2009fundamentals}. 
Determining absolute configuration and enantiomeric excess is therefore essential.
Conventional chiroptical techniques rely on the spin angular momentum (SAM) of light, i.e. its polarization \cite{berova2000circular, barron2009molecular}, and provide enantiomeric sensitivity in isotropic ensembles, yet they rarely yield direct atomic-scale structural information.

Standard x-ray diffraction (XRD) as a structural probe is generally insensitive to inversion-symmetry breaking in the non-resonant regime, irrespective of whether the sample is crystalline \cite{parsons2017determination}.
For linear non-resonant elastic x-ray scattering, polarization only enters as the Lorentz-polarization prefactor~\cite{als2011elements} and does not couple to the scalar charge density.
This results in Friedel's law: diffracted intensities at $\pm \bm q$ momentum transfer are equal \cite{friedel1913symetries, parsons2017determination}, so enantiomers cannot be distinguished and their absolute configuration cannot be determined~\cite{hooft2008determination}.
Close to an electronic resonance, anomalous scattering introduces complex atomic scattering factors, allowing Bijvoet pairs to differ in intensity.
This Bijvoet difference, together with anomalous dispersion, has been used for absolute structure determination \cite{parsons2017determination, flack2008use} in crystals. Recently a time-resolved scattering approach has been proposed to probe charge migration in a chiral molecule~\cite{giri2021imaging}.

\begin{figure*}[t!]
    \centering
    \includegraphics[width=0.99\textwidth]{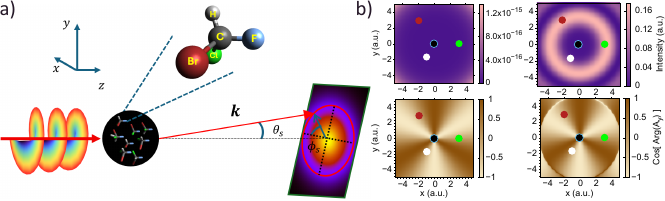}
    \caption{ 
a) Scattering geometry for twisted x-ray diffraction from an axially oriented molecular ensemble. Inset: CHBrClF molecule (C–F along $+z$). 
b) Intensity (top) and phase profiles (bottom) in the $z=0$ plane for OAM beams with $\omega_\text{in}=9.25$ keV, $m = +4$, $\Lambda=+1$ and cone angles $\theta_k = 0.1\degree$ (left) and $\theta_k = 30 \degree$(right). 
Atomic positions shown (C: black; H: white; Br: maroon; Cl: green; F: blue) for a single oriented molecule with carbon at the beam center for some arbitrary axial rotation. 
Only the phase of $A_y$ is shown. 
These two cases sample regimes with markedly different paraxial-approximation accuracy for the twisted field–matter interaction.
    }
    \label{fig:intro_schematics}
\end{figure*}

The increasing availability of spatially structured x-rays enables new characterization methods \cite{forbes2021structured} with x-rays carrying an Orbital Angular Momentum (OAM) attracting particular interest \cite{sasaki2008proposal, ye2019probing, huang2021generating, woods2021switchable, yan2023self, mccarter2024generation}.
In addition to their SAM, these beams, also known as twisted or vortex beams, possess a spatially structured wavefront that has a topological charge \cite{allen1999iv, torres2011twisted}.
This usually manifests as a rotating phase in the field components (see Fig.~\ref{fig:intro_schematics}).
Twisted beams have been widely exploited in the optical domain \cite{shen2019optical, nazirkar2025manipulating} and are emerging in x-ray applications including ptychography \cite{pancaldi2024high} and spectroscopy \cite{rouxel2022hard}. 

Since the OAM light-matter interaction provides a field pseudoscalar~\cite{green2023optical}, it opens new routes to probe chirality-sensitive processes.
Advantageously, OAM beams with their large set of controllable parameters offers additional paths for signal optimization.
While twisted x-ray diffraction from chiral nanocrystals have been proposed \cite{nazirkar2024coherent}, it remains unclear whether such effects persist in x-ray diffraction from isotropic chiral media.

In the paraxial limit, the twisted wavefront and polarization decouple, and symmetry arguments enforce Friedel's law for ground state molecules.
Yong \textit{et al.} \cite{yong2022direct} have shown that time-resolved x-ray scattering from a single photoexcited molecule at the OAM beam center can exhibit a dichroic response, interpreted as a signature of conical-intersection passage,
with related extensions to electron-vortex scattering \cite{wu2025diffractive}.
Molecular chirality measurements in liquids and gases are inherently ensemble averaged.
Single-molecule x-ray diffraction, which can avoid such averaging, remains experimentally challenging \cite{neutze2000potential, chapman2019x}.
This raises the question of whether the increased field gradients of tightly focused beams can reveal chiral structures under rotational and focal averaging.
In this regime which lies beyond the paraxial approximation, the OAM and SAM are coupled and only the total angular momentum is well defined.
Furthermore, Forbes et al. \cite{forbes2024topological} have shown that coupling between the longitudinal and transverse components of an OAM field can play a significant role in light-matter interactions, with this coupling becoming increasingly pronounced beyond the paraxial regime.

In this work, we establish under which conditions a chirality-sensitive dichroic signal can exist under nonresonant elastic scattering.
We first prove using an irreducible expansion of the charge density that randomly oriented molecules cannot yield a difference signal between enantiomers.
We then demonstrate through simulations that a nonvanishing dichroic signal can arise in the case of a single oriented molecule.
We determine the conditions under which a dichroic scattering response can arise in an ensemble of chiral molecules, and show that these conditions are more restrictive than previously anticipated due to focal averaging effects and ensemble response. 
Atomic units are used throughout unless stated otherwise.

\section{Scattering from an isotropic ensemble}

\begin{figure*}
\includegraphics[width=0.95\textwidth]{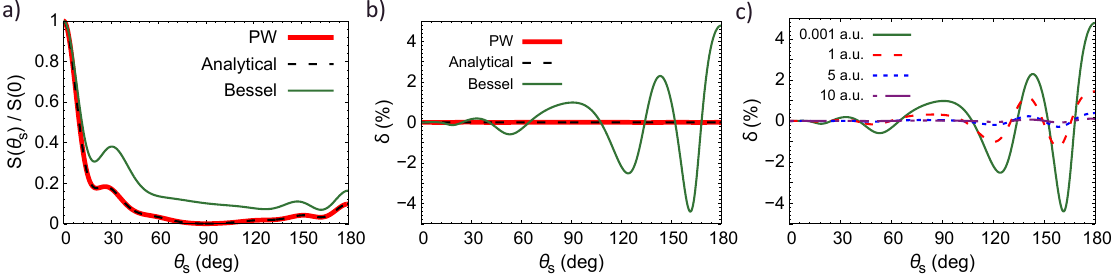}
\caption{
a) Scattering yields for a single molecule at the beam center ($\sigma_b = 0.001$ a.u.) after axial averaging: linearly polarized plane wave (PW), Debye-scattering analog for axial averaging (Analytical), and OAM Bessel beam (Bessel) with $\theta_k = 30\degree$, $m= 4$, and $\Lambda = 1$. 
b) Dissymmetry factor $\delta$ (percentage difference in yield) between enantiomers. 
c) $\delta(\theta_s)$ for different rms positions $\sigma_b$ with respect to the beam center, revealing focal averaging effects. 
$\phi_s = 90\degree$ for all cases. The single molecule scattering calculations from plane wave and Bessel beams are obtained from averaging over $10^8$ and $10^9$ realizations, respectively.
}
\label{fig:single_molecule}
\end{figure*}

For a monochromatic incident field of photon energy $\omega_\text{in}$, the vector potential is written as
\begin{equation} \label{harmonicfield}
\bm A_\text{in}(\bm r,t) = \bm A(\bm r) e^{-i \omega_\text{in} t} + \bm A^*(\bm r) e^{i \omega_\text{in} t},
\end{equation}
where $\bm r$ and $t$ denote position and time, respectively.
The XRD signal $s$ for a beam with an arbitrary spatial structure $A(\bm r)$ is given by \cite{bennett2018monitoring}:
\begin{multline}
s \propto \int d\bm r d\bm r' \ \langle\sigma(\bm r)\sigma^*(\bm r') \rangle_\Omega\\
\times (\bm A(\bm r)\cdot \bm \epsilon^*)
(\bm A^*(\bm r')\cdot \bm \epsilon)
e^{-i\bm k_s\cdot (\bm r - \bm r')}
\end{multline}
\noindent where $\sigma(\bm r)$ is the total matter charge density, $\bm \epsilon$ and $\bm k_s$ are the scattered field polarization and wavevector, respectively.
$\langle ...\rangle_\Omega$ indicates a rotational averaging performed over the molecular orientations.

The two enantiomers of a chiral compound are transformed into each other by a parity operation.
Thus, a signal capable of distinguishing them needs to involve a pseudoscalar \cite{barron2009molecular}.
In many dichroic measurements, probing the difference signal between the $\Delta$ and $\Lambda$ enantiomers with a given incident field can be mimicked by making a differential measure with complementary chiral fields (usually switching polarization between left and right) on a given enantiomer.
Outside the paraxial regime, switching the chirality of the twisted fields becomes non-trivial, therefore, we examine the difference yield between the enantiomers.

First, we show that a randomly oriented molecule produces no dichroic signal, irrespective of the incident field's spatial structure.
The charge densities for the opposite enantiomers are noted $\sigma^\Delta(\bm r)$ and $\sigma^\Lambda(\bm r)$.
Their spherical harmonics expansion is given by:
\begin{align}
\sigma^\Delta(\bm r) &= \sum_{lm} \sigma_{lm}(r) \ Y_{lm}(\hat r)\\
\sigma^\Lambda(\bm r) &=  \sum_{lm} (-1)^l \sigma_{lm}(r) \ Y_{lm}(\hat r)
\end{align}
\noindent where we have used that the charge densities for opposite enantiomers transform into one another by a parity operation up to a rotation ($\mathcal P \sigma^\Delta(\bm r) = \sigma^\Lambda(\bm r')$), and that the spherical harmonics pick up a $(-1)^l$ factor upon parity: $\mathcal P Y_{lm}(\hat r) = (-1)^l Y_{lm}(\hat r)$. 

The rotationally averaged two-point correlation function of the charge density is expanded as
\begin{multline}
\langle \sigma(\bm r)\sigma^*(\bm r') \rangle_\Omega =
\langle \sum_{lm} \sum_{l'm'}
\sigma_{lm}(r)\sigma_{l'm'}^*(r')\\ 
\times Y_{lm}(\bm{\hat r}) Y_{l'm'}^*(\bm{\hat r'})\rangle_{\Omega}.
\end{multline}

The differential matter correlation function is then given by:
\begin{multline}
\langle\sigma^\Delta(\bm r)\sigma^{\Delta*}(\bm r') \rangle_\Omega
- \langle\sigma^\Lambda(\bm r)\sigma^{\Lambda*}(\bm r') \rangle_\Omega = \\
\sum_{lm}\sum_{l'm'} \sigma_{lm}(r) \sigma_{l'm'}^*(r') \langle Y_{lm}(\hat r) Y_{l'm'}^*(\hat r')\rangle_\Omega\\
- \sum_{lm}\sum_{l'm'} (-1)^l (-1)^{l'} \sigma_{lm}(r) \sigma_{l'm'}^*(r') \langle Y_{lm}(\hat r) Y_{l'm'}^*(\hat r')\rangle_\Omega\\
=\sum_{lm}\sum_{l'm'}  (1 - (-1)^{l+l'} ) \sigma_{lm}(r) \sigma_{l'm'}^*(r') \langle Y_{lm}(\hat r) Y_{l'm'}^*(\hat r')\rangle_\Omega
\end{multline}

The rotational average of the spherical harmonics can be explicitly carried out:
\begin{multline}
\langle Y_{lm}(\hat r) Y_{l'm'}^*(\hat r')\rangle_\Omega = \sum_{m_1m_2}Y_{lm_1}(\hat r) Y_{l'm_2}^*(\hat r')\\
\times \Big( \int \frac{d\Omega}{8\pi^2} \mathcal D^l_{mm_1}(\Omega)\mathcal D^{l'*}_{m'm_2}(\Omega) \Big) 
\end{multline}

\noindent where the Wigner $\mathcal D$ matrices obey the following orthogonality relationship:
\begin{equation}
\int \frac{d\Omega}{8\pi^2} \mathcal D^l_{mm_1}(\Omega)\mathcal D^{l'*}_{m'm_2}(\Omega) 
= \frac{1}{2l+1}\delta_{ll'}\delta_{mm'}\delta_{m_1m_2}
\end{equation}

The enantiomer difference of the two-point charge correlation functions becomes:
\begin{multline}
\langle\sigma^\Delta(\bm r)\sigma^\Delta(\bm r') \rangle_\Omega
- \langle\sigma^\Lambda(\bm r)\sigma^\Lambda(\bm r') \rangle_\Omega = \\
\sum_{lm}\sum_{l'm'}\sum_{m_1m_2}  (1 - (-1)^{l+l'}) \ 
\sigma_{lm}(r) \sigma_{l'm'}^*(r') \\
\times \frac{1}{2l+1}\delta_{ll'}\delta_{mm'}\delta_{m_1m_2}
Y_{lm_1}(\hat r) Y_{l'm_2}^*(\hat r')
= 0
\end{multline}

This cancellation depends only on the matter terms and is independent of the incident beam spatial profile.

\begin{figure*}
\resizebox{160mm}{!}{\includegraphics[]{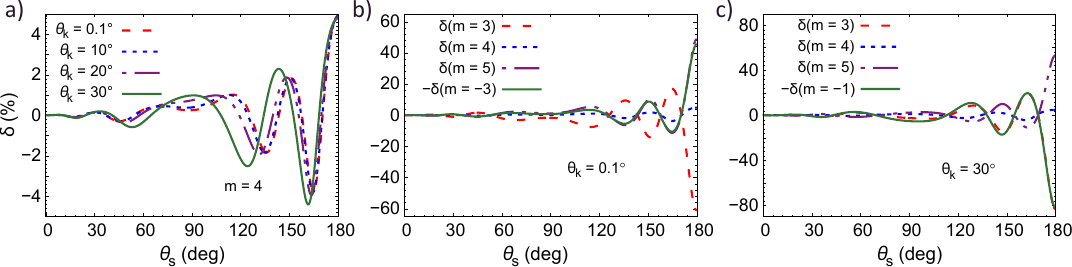}}
\caption{ Dissymmetry factor $\delta$ vs Bessel beam parameters. a) $\theta_k$ dependence for fixed $m = 4$. b) $m$-dependence for $\theta_k=0.1\degree$. c) $m$-dependence for $\theta_k = 30\degree$. In all cases, $\Lambda = 1$, $\phi_s = 90\degree$, $\sigma_b = 0.001$ a.u.}
\label{fig:singlemol_besselparamdep}
\end{figure*}

\section{Scattering from an oriented sample}

To retain parity-odd contributions, we consider axially oriented molecules, i.e. molecules oriented along a single axis and not full three-dimensional orientation. 
In this case, axial (cylindrical) averaging replaces full rotational averaging.
For plane wave x-ray diffraction, similar to the Debye scattering equation for full rotational averaging, an analogous expression for oriented molecules was derived previously ~\cite{inouye1993structure, ross2020extending}. 
The scaled scattering yield for a plane wave incident on a cylindrically averaged group of atoms is:
\begin{equation} \label{eq:Debye_axial_avg}
\begin{split}
\frac{S(\bm q)}{S_e} = \sum_{ij} & f_i(q) f_j(q) \cos(q_z r_{ijz}) \\
& \times J_0\Big(r_{ijxy} \sqrt{q_x^2 + q_y^2}\Big),
\end{split}
\end{equation}
\noindent where $r_{ijxy} = \sqrt{(r_{ix}- r_{jx})^2 + (r_{iy}- r_{jy})^2}$, $r_{ijz} = |r_{iz}- r_{jz}|$, $f_i(q)$ is the atomic form factor of the $i$th atom, $\boldsymbol{q} = \boldsymbol{k}_\text{in} -\boldsymbol{k}_s$, $S(\bm q)$ is the scattering yield, and $S_e$ is the Thomson scattering yield from a free electron. The quantities $\boldsymbol{k}_\text{in}$ and $\boldsymbol{k}_s$ denote the incident photon momentum of the plane wave and the scattered photon momentum, respectively.

In this work, we simulate the elastic scattering response numerically using a monochromatic OAM Bessel beam ~\cite{boning2018above, matula2013_Besselbeam}. 
The spatial part $\bm{A}(\bm{r})$ of the twisted Bessel field from B{\"o}ning et al.~\cite{boning2018above} is expressed here as $\bm{A}(\bm{r}) =  \bm{g}(\rho,\phi)~e^{i k_z z}$.
The position vector $\bm{r}(\rho,\phi,z)$ is given in cylindrical coordinates where $\rho$, $\phi$ and $z$ denote the radial, azimuthal and axial coordinate, respectively. The vector $\bm{g}$ is defined as
\begin{equation} \label{eq:g_x}
    \begin{split}
        g_x =  & \frac{1}{2} A_0 \sqrt{\frac{k_{\perp} } {4\pi} }  \bigg\{ c_{-1} J_{m+1}(k_{\perp} \rho) e^{i(m+1)\phi} \\
        &+ c_{+1} J_{m-1}(k_{\perp} \rho) e^{i(m-1)\phi} \bigg\},
    \end{split}
\end{equation}
\begin{equation} \label{eq:g_y}
    \begin{split}
        g_y =  & \frac{1}{2i} A_0 \sqrt{\frac{k_{\perp} } {4\pi} }  \bigg\{ c_{-1} J_{m+1}(k_{\perp} \rho) e^{i(m+1)\phi } \\
        &- c_{+1} J_{m-1}(k_{\perp} \rho) e^{i(m-1)\phi} \bigg\},
    \end{split}
\end{equation}
\begin{equation} \label{eq:g_z}
        g_z =  \frac{1}{2i} A_0 \sqrt{\frac{k_{\perp} } {2\pi} } c_{0} J_{m}(k_{\perp} \rho) e^{im\phi}.
\end{equation}
\noindent Here $k_\perp = (\omega_\text{in}/c) \sin \theta_k$ and $k_z = (\omega_\text{in}/c) \cos \theta_k $ are the perpendicular and parallel components of the linear momentum of the Bessel beam where $c$ is the speed of light in vacuum and $\theta_k$ is the OAM beam cone opening angle. 
The coefficients are $c_{\pm 1} = (1 \pm \Lambda \cos \theta_k)/$2 and $c_0 = \Lambda \sin \theta_k / \sqrt{2}$.
The quantities $A_0$, $J_m$, $m$, and $\Lambda$ are the field amplitude, Bessel function of the first kind, projection of the total angular momentum (TAM) along the propagation direction, and beam helicity, respectively.
The cone angle $\theta_k$ determines the radial spacing of the Bessel zeros and thus controls the beam's transverse size.

The above field $\bm A(\bm r)$ [Eqs.~(\ref{eq:g_x})-(\ref{eq:g_z})] 
is an exact solution of the Helmholtz equation and an eigenvector of the total angular momentum projection along beam-axis $J_z$ with eigenvalue $m$ \cite{matula2013_Besselbeam}.
Although the intensity profile for $|m|$ and $-|m|$ are not identical, in the paraxial limit the intensity profiles for $|m|$ and $2-|m|$ become approximately equal and the fields become approximately complementary (see Appendix~\ref{Appendix_Intensity_twisted} for the intensity expression).
In that paraxial limit, SAM and OAM decouple to yield well-defined corresponding values $\Lambda$ and ($m-\Lambda$) respectively and the component $A_z(\bm r) \approx 0$.
For $\theta_k \rightarrow 0$, with $m=\Lambda=\pm1$, the field reduces to circularly polarized field.

\subsection{Computational approach}

\begin{figure}
\resizebox{77mm}{!}{\includegraphics[]{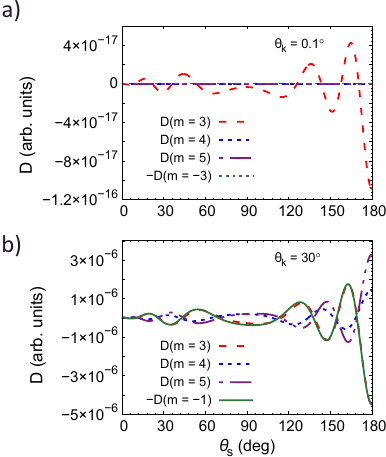}}
\caption{Enantiomeric raw difference in scattered yield vs Bessel beam parameters. a) $m$-dependence for $\theta_k=0.1\degree$. b) $m$-dependence for $\theta_k = 30\degree$. Here, $\Lambda = 1$, $\phi_s = 90\degree$, $\sigma_b = 0.001$ a.u.
}
\label{fig:singlmol_diffsig_besselparam}
\end{figure}

We investigate twisted x-ray diffraction from chiral molecules using the independent atom model. 
We retain the field's spatial variation across atomic centers while neglecting intra-atomic field variations. 
For a single molecule with scattered photon momentum $\bm k_s$ and polarization $\bm  \epsilon$, the scattering amplitude is (see Appendix.~\ref{Appendix_TWXS_amplitude_derivation})
\begin{equation} \label{Scattering_amplitude_molecule}
    F_1 = \sum_{a}^{n_a} f_a(\boldsymbol{Q}) ~ e^{i \boldsymbol{Q}\cdot \boldsymbol{R}_a}  ~ \bm \epsilon^* \cdot \boldsymbol{g}(\rho_a, \phi_a).
\end{equation}
\noindent where $\boldsymbol{Q} = \boldsymbol{k}_z -\boldsymbol{k}_s$ and $\boldsymbol{k}_z = k_z \hat{z}$. 
The quantities $f_a(\boldsymbol{Q})$ and $\boldsymbol{R}_a$ are the atomic form factor and position of the atom $a$, respectively. $n_a$ denotes the number of atoms in a single molecule. 

We simulate twisted x-ray diffraction on an archetypal chiral molecule, bromochlorofluoromethane (CHBrClF) with the C-F bond fixed along the $z$-direction (Fig.~\ref{fig:intro_schematics}). 
The molecular coordinates were obtained by geometry optimization using classical force field in Avogadro \cite{Avogadro_reference}.
We compute scattering for both a single molecule and for molecular ensembles.

For the single-molecule simulations, we evaluate the scattering yield $S = \abs{F_1}^2$ for a uniform random axial rotation around the C-F bond over the interval $[0,2\pi]$, and for an impact parameter drawn from a zero-mean Gaussian distribution with root mean square (rms) value $\sigma_b$ for each cartesian coordinate of the carbon atom in the molecule.
Then the scattering yield is averaged over a billion realizations.
The parameter $\sigma_b$ models focal-volume-limited interaction and tolerance in the impact parameter. 
Physically $\sigma_b$ characterizes the average molecule distance from the beam centre.

For ensembles, the scattering yield is the absolute square of the coherent sum of individual molecular scattering amplitudes (each molecule generated as above). The total scattering amplitude from an ensemble of $n_M$ molecules is given by
\begin{equation} \label{Scattering_amplitude_total}
    F = \sum_{b=1}^{n_M} F_{b}.
\end{equation}
Here $F_{b}$ is the scattering amplitude from molecule $b$. For a given scattered momentum direction ($\theta_s$, $\phi_s$), the scattered photon momentum is given by $\boldsymbol{k}_s = k_s (\sin \theta_s \cos \phi_s,~\sin \theta_s \sin \phi_s,~\cos \theta_s)$ and two allowed choices for outgoing photon polarization are
\begin{equation} 
\begin{split} \label{polarization_choices}
     &\boldsymbol{\epsilon}_1 = (\cos \theta_s \cos \phi_s,~\cos \theta_s \sin \phi_s,~ -\sin \theta_s), \\
     &\boldsymbol{\epsilon}_2 = (-\sin\phi_s,~\cos \phi_s,~0).
\end{split}
\end{equation}
The total scattered yield when the outgoing polarization is not measured is given by
\begin{equation} \label{Scattering_yield}
    S = \sum_{\bm \epsilon} |F|^2 .
\end{equation}
\noindent where $\sum_{\bm \epsilon}$ denotes a sum over polarization components.
The total scattered yield is averaged over many ensemble realizations for $n_M$ molecules. The ensemble averaged yield $\expval{S}$ is proportional to the scattering signal measured in an experiment.

Ensemble results reported in the main text are obtained from averaging over 30 million realizations. A discussion of the numerical convergence of the calculated results with increasing number of ensemble realizations is provided in Appendix.~\ref{Appendix_convergence}.
All reported yields are summed over the two outgoing photon polarizations (i.e. the scattered photon polarization is not measured).
Given the scattering yields $S_{\Delta}$ and $S_{\Lambda}$ for the $\Delta$ and $\Lambda$ enantiomers, respectively, the dissymmetry factor $\delta$ is defined as
\begin{equation} \label{Dichroic_signal}
    \delta = \frac{\expval{S_{\Delta}} - \expval{S_{\Lambda}} }{  0.5 \big(\expval{S_{\Delta}} + \expval{S_{\Lambda}}\big) } \times 100.
\end{equation}
The raw difference in scattering yield between the enantiomers $D$ is defined as 
\begin{equation} \label{eq:Raw_diff_yield} 
    D = \langle S_{\Delta}\rangle - \langle S_\Lambda\rangle.
\end{equation}

\begin{figure}
\resizebox{72mm}{!}{\includegraphics[]{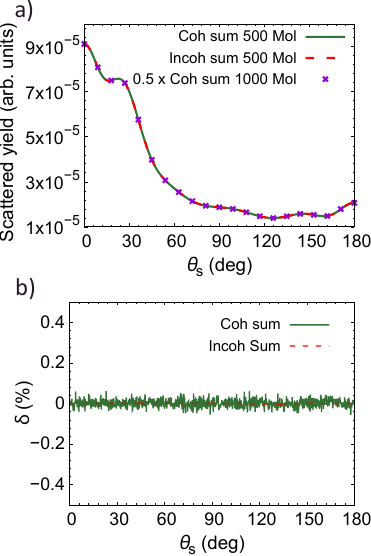}}
\caption{ 
a) Scattering yields for ensembles of 500 and 1000 molecules. “Coh sum”: $|\sum_i F_i|^2$, “Incoh sum”: $\sum_i |F_i|^2$.
b) $\delta$ for the 500-molecule ensemble. 
Here, $\sigma_b = 1000$ a.u. ($\sim 53$ nm). 
The scattering calculations for a molecular ensemble are obtained from averaging over 30 million realizations. The other parameters are identical to Fig.~\ref{fig:single_molecule}.
}
\label{fig:many_mol}
\end{figure}

\subsection{Results and Discussion}
Figure~\ref{fig:single_molecule}a shows the scattering yield of a single $\Delta$ molecule located at the beam center. 
The plane-wave result (solid red curve, PW) is for a linearly polarized incident field for in-plane scattering ($\phi_s = 90\degree$).
The yield obtained from the analytical expression in Eq.~(\ref{eq:Debye_axial_avg}) is also displayed.
Its agreement with our stochastic plane-wave simulation provides a benchmark.
Notably, the plane-wave yield vanishes at $\theta_s = 90\degree$, whereas the twisted yield does not. 
For a linearly polarized plane wave, the polarization factor $\abs{ \bm \epsilon \cdot \bm \epsilon_\text{in}}^2$ vanishes at $\theta_s=90\degree$ for in-plane scattering ($\phi_s = 90\degree$), which is not the case for the twisted Bessel field [Eq.(~\ref{Scattering_amplitude_molecule})].
Furthermore, the plane-wave scattering yield depends on the azimuthal angle $\phi_s$ due to this polarization factor and not due to the underlying molecular structure.
In contrast, the twisted beam yields show no azimuthal angle dependence.

Figure~\ref{fig:single_molecule}b shows the dissymmetry factor $\delta$ [Eq.~(\ref{Dichroic_signal})]. 
For conventional plane wave x-ray diffraction, $\delta$ vanishes for all scattering angles, even for a perfectly oriented chiral molecule. In contrast, a twisted x-ray field yields a clear non-zero dichroic signal when the molecule is centered in the beam. 
The twisted field introduces intensity and phase gradients across the molecule (Fig.~\ref{fig:intro_schematics}b), leading to different phases and probabilities for the scattered wave from each atom in the molecule.
The resulting interference pattern remains chiral-sensitive after axial averaging. 
The dissymmetry factor is zero in the forward direction and maximal in backward elastic scattering. 
In the paraxial limit, the forward cancellation can be shown analytically: complementary twisted beams produce equal and opposite contributions from atomic pairs (see Appendix.~\ref{Appendix_paraxial_approx}).
Our result demonstrates that this vanishing persists well beyond the paraxial regime, confirms trends implicit in Ref.~\cite{nazirkar2024coherent}, and clarifies their origin.
While for Fig.~\ref{fig:single_molecule}b, the largest $\delta$ is for backward scattering, it should be noted from Fig.~\ref{fig:single_molecule}a that the backward scattering photon yield is about $16\%$ of the forward scattering yield.
 
Figure~\ref{fig:single_molecule}c shows the dependence of the $\delta$ on the rms position $\sigma_b$. 
Relaxing the restriction that the molecule is at the beam center introduces both axial and positional averaging, leading to a rapid suppression of the dichroic signal. 
As the molecule is located further from the beam center, the field gradient across its atoms decreases. 
For large displacements, the molecule effectively experiences a plane wave, whose dichroic response does not survive axial averaging. 
Focal averaging plays a decisive role in the survival of dichroism in disordered systems illuminated by twisted beams.

Figure~\ref{fig:singlemol_besselparamdep} reveals the dependence of $\delta$ on the Bessel beam parameters (TAM $m$ and cone angle $\theta_k$) for a molecule near the beam center.
Increasing $\theta_k$ (Fig.~\ref{fig:singlemol_besselparamdep}a) causes both intensity and phase gradients across the molecules, yielding a moderate increase in $\delta$.

An intuitive argument can be made for the non-monotonic dependence of the dichroic signal on $m$ (see Fig.~\ref{fig:singlemol_besselparamdep}b) as a function of the molecular geometry: for the chosen system (central chirality and molecule oriented along one of the bonds), the atoms are approximately organized in a three-fold arrangement, which matches the periodicity of the phase rotation of an $m=4$ beam (see Fig.~\ref{fig:intro_schematics}b).
Thus, each atom experiences almost the same relative phase of the OAM beam and the dichroic signal is strongly reduced.
Alternatively, other $m$ values that cause larger phase gradients between different atomic sites across the molecules will increase the effect of the twisted wavefronts on the scattering signal. 
Since beams with $J_z$ values $|m|$ and $-|m|$ have different intensity profiles, their dichroic responses are not identical.
However, in the paraxial approximation, replacing $m$ by its complementary value $2-|m|$ approximately mimics the effect on the scattering yield from mirror inversion of the molecule.
The calculations (Figs.~\ref{fig:singlemol_besselparamdep}b and ~\ref{fig:singlemol_besselparamdep}c)  show that this approximation is most accurate for backward elastic scattering and that $\delta$ is maximized there. In addition, it should be noted that the m-dependence of dissymmetry factor would be sensitive to the specific molecule and the chosen orientation axis.

In structured light studies, often the most interesting effects may manifest in the region where the absolute scattered yield is low. Figure~\ref{fig:singlmol_diffsig_besselparam} shows the scattering angle dependence of the enantiomeric raw difference in the photon yield [Eq.~(\ref{eq:Raw_diff_yield})] instead of $\delta$ which is normalized. 
This illustrates the relative size of the observable difference signal for different $m$ values. Figure~\ref{fig:singlmol_diffsig_besselparam} further reveals that the  absolute enantiomeric difference yield also reaches a maximum for the same $m$ values and near the same scattering angles which exhibit the highest dissymmetry factors. The low yield for the curves for $\theta_k = 0.1 \degree$ (Fig.~\ref{fig:singlmol_diffsig_besselparam}a) relative to $\theta_k = 30\degree$ (Fig.~\ref{fig:singlmol_diffsig_besselparam}b) is due to the low intensity near the beam centre where the molecule is located (refer Fig.~\ref{fig:intro_schematics}b). 

Figure~\ref{fig:many_mol} considers an ensemble with rms position $\sigma_b \approx 53$ nm (1000 a.u.), comparable to achievable x-ray focal size. 
Figure~\ref{fig:many_mol}a shows the scattering yield for ensembles of 500 and 1000 molecules. The chosen $\sigma_b$ and the number of molecules roughly correspond to an ideal gas at 300 K with pressures of about 0.14 and 0.28 atm, respectively.
The coherent sum is $|\sum_i F_i|^2$, whereas the incoherent sum is  $\sum_i|F_i|^2$.
Note that $|F_1|^2$ [Eq.~(\ref{Scattering_amplitude_molecule})] already includes coherent (intra-molecular) interference among atomic amplitudes, which generates the dichroism. 
The agreement between coherent and incoherent sums indicates the absence of inter-molecular interference upon ensemble averaging. The random positions and random axial orientations of the molecules ensure that there exists no fixed phase difference between the incident fields experienced by two molecules. This gives rise to scattered fields from two molecules with no fixed phase difference.
Consequently, the total twisted scattered yield scales linearly with number of molecules, analogous to conventional XRD.

Figure~\ref{fig:many_mol}b shows that the ensemble dissymmetry factor $\delta$ effectively vanishes for all scattering angles. 
The dichroic signal arises predominantly from molecules near the beam axis.
Molecules far from the center effectively experience a locally plane-wave incident field and do not encounter the twisted spatial structure. 
The limited number of molecules at the center, together with the reduced on-axis intensity, render the difference yield to total yield ratio negligible. 
Small deviations from zero reflect finite statistical sampling. 
The dissymmetry factor $\delta$ computed from only the incoherent sum is closer to zero than that from the coherent sum which exhibits small high-frequency oscillations, indicating that the dominant statistical noise originates from intermolecular interference terms which do not cancel for a finite number of realizations.
These results demonstrate that focal averaging, i.e. the inability to confine molecules close to the beam axis, critically suppresses chiral sensitivity in disordered ensembles.

For polarization-sensitive detection, the exact qualitative shape of the dissymmetry factor $\delta$ vs $\theta_s$ depends on the scattered photon polarization. However, we find that for large $\sigma_b$ (Figs.~\ref{fig:single_molecule}c and \ref{fig:many_mol}b) where focal averaging suppresses the dissymmetry factor obtained from sum over polarizations, the dissymmetry factor is also diminished for each of the two choices of scattered photon polarization [Eq.~(\ref{polarization_choices})]. Our calculations suggest that for experimentally achievable beam focus sizes, polarization selectivity alone cannot substantially overcome focal averaging effects.

\section{Conclusion and Summary}
In this work, we theoretically investigated x-ray scattering with twisted Bessel beams to probe chiral molecules with simulations presented for a simple chiral molecular system CHBrClF.
We prove that irrespective of the properties of the incident beam, upon full rotational averaging, there exists no difference in the nonresonant elastic scattering signal between enantiomers. 
However, for an oriented molecule, there exists a difference signal between the enantiomers which depends on the parameters of the Bessel beam. In addition, the dichroic signal from an oriented single molecule diminishes due to focal averaging as the rms position from the beam centre increases.
Scattering calculations from an ensemble of oriented chiral molecules highlight the pivotal role focal averaging effects can play in diminishing an observable dichroic signal independent of any temperature effects on orientation. 
The results also strongly suggest that twisted x-ray scattering is likely to be effective in probing chirality in crystalline samples where these effects are reduced. Other potential yet challenging pathways to minimize focal averaging effects in samples include considering nanoclusters whose chiral features manifest over large length scales or alternatively resorting to trapping molecules over a narrow focal volume.

\section*{Acknowledgements}
A.V. benefited from prior discussions and early modelling of twisted Bessel beams with F. Robicheaux. This work was supported by the U.S. Department of Energy, Office of Basic Energy Sciences, Division of Chemical Sciences, Geosciences, and Biosciences through Argonne National Laboratory. Argonne is a U.S. Department of Energy laboratory managed by UChicago Argonne, LLC, under Contract No. DE-AC02-06CH11357. We gratefully acknowledge the computing resources provided on Improv, a high-performance computing cluster operated by the Laboratory Computing Resource Center at Argonne National Laboratory.

\section*{Data availability}
The data presented in the figures will be available upon reasonable request.

\appendix

\section{Twisted x-ray scattering amplitude} \label{Appendix_TWXS_amplitude_derivation}

The elastic x-ray scattering signal $s$, from a single electron is proportional to the absolute square of the transition matrix element involving $\hat{\bm A}^2$ interaction term~\cite{Sakurai_adv},
\begin{equation} \label{Scattering_yield_1electron}
    s \propto   \abs{F_\text{electron}}^2
\end{equation}
where
\begin{equation} \label{Scattering_amplitude_1electron}
    F_\text{electron} = \bra{\psi} e^{-i \bm k_s\cdot \boldsymbol{r}} \bm \epsilon^* \cdot \bm A(\boldsymbol{r}) \ket{\psi}.
\end{equation}
Here, $\ket{\psi}$ describes the state of the electron, $\boldsymbol{k}_s$ is the scattered photon momentum, and $\boldsymbol{\epsilon}$ is the scattered photon polarization, respectively.
For a linearly polarized incident plane wave with incident momentum $\bm{k_0}$ and polarization $\bm \epsilon_\text{in}$, the scattering amplitude from a bound electron is \mbox{$F_\text{pw} = \bm \epsilon^* \cdot \bm \epsilon_\text{in} \bra{\psi} e^{i (\bm{k}_0-\bm{k}_s)\cdot \bm{r}} \ket{\psi}$ }.

For a single molecule, the total scattering amplitude is the sum of individual scattering amplitudes from every electron in every atom.
\begin{equation} \label{Scattering_amplitude_initial}
    F_1 = \bra{\Psi} \sum_{a}^{n_a} \sum_j^{n_e(a)} e^{-i \boldsymbol{k}_s\cdot (\boldsymbol{R}_a + \boldsymbol{r}_{aj}}) \bm \epsilon^* \cdot \bm A(\boldsymbol{r}) \ket{\Psi}.
\end{equation}
The index $a$ iterates over all atoms in a given molecule and the index $j$ iterates over the number of the electrons ($n_e$) in atom $a$. The quantity $\boldsymbol{R}_a$ denotes the position of the nucleus of atom $a$ and $\boldsymbol{r}_{aj}$ is the position vector for electron $j$ from the nucleus of atom $a$. The quantity $\ket{\Psi}$ describes the ground state of the molecule. In the independent atom model, 
the total molecular wavefunction reduces to a product of individual atomic wavefunctions. 
Therefore the above expression becomes
\begin{multline} 
\label{Scattering_amplitude_general}
F_1 = 
\sum_{a}^{n_a} e^{-i \bm{k}_s\cdot \bm{R}_a}  \bra{\psi_{a}} 
\sum_j^{n_e(a)} e^{-i \bm{k}_s\cdot \bm{ r}_{aj}}\\
\times\bm \epsilon^* \cdot \bm A(\bm{R}_a + \bm{r}_{aj} ) \ket{\psi_{a}},
\end{multline}
where $\ket{\psi_{a}}$ refers to the electronic wavefunction of atom $a$.

In this work, we examine elastic off-resonant scattering from an incident twisted Bessel x-ray field $\boldsymbol{A}_\text{in}$ propagating along the $z$-direction. 
The exact expression for a twisted Bessel field has been derived previously by Matula et al. ~\cite{matula2013_Besselbeam} and B\"oning et al.~\cite{boning2018above}.
The cartesian components of the twisted field~\cite{boning2018above} are 
\begin{widetext}
\begin{equation} \label{A_x}
    \begin{split}
        A_{\text{in},x}(\boldsymbol{r},t) =  A_0 \sqrt{\frac{k_{\perp} } {4\pi} }      \bigg\{ & c_{-1} J_{m+1}(k_{\perp} \rho)   ~\cos{ \big[(m+1)\phi + k_z z -\omega_\text{in} t \big] }  \\
         & + c_{+1} J_{m-1}(k_{\perp} \rho)~\cos{\big[(m-1)\phi + k_z z -\omega_\text{in} t \big] }
        \bigg\},
    \end{split}
\end{equation}
\begin{equation} \label{A_y}
    \begin{split}
        A_{\text{in},y}(\boldsymbol{r},t) = A_0 \sqrt{\frac{k_{\perp} } {4\pi} }  \bigg\{ & c_{-1} J_{m+1}(k_{\perp} \rho) \sin{ \big[(m+1)\phi + k_z z -\omega_\text{in} t \big]} \\
        & - c_{+1} J_{m-1}(k_{\perp} \rho) \sin{ \big[(m-1)\phi + k_z z -\omega_\text{in} t\big] }  \bigg\},
      \end{split}
\end{equation}

\begin{equation} \label{A_z}
A_{\text{in},z}(\boldsymbol{r},t) = A_0 \sqrt{\frac{k_{\perp} } {2\pi} } c_{0} J_{m}(k_{\perp} \rho) \sin{ (m\phi + k_z z -\omega_\text{in} t \big) }.
\end{equation}
\end{widetext}
where
\begin{equation} \label{eq:defn_c_coeff}
c_{\pm1} = \frac{1}{2}(1 \pm \Lambda \cos \theta_k ),~~
c_{0} = \frac{\Lambda}{\sqrt{2}} \sin \theta_k.
\end{equation}
Here $A_0$, $m$, $\Lambda$, and $\theta_k$ are the amplitude of the incident field's vector potential, projection of the angular momentum along the propagation direction, beam helicity, and cone opening angle, respectively. The position vector $\boldsymbol{r}(\rho,\phi,z)$ is expressed in cylindrical coordinates with $\rho$, $\phi$ and $z$ denoting the radial coordinate, azimuthal angle and z coordinate, respectively. The quantities $k_\perp = (\omega_\text{in}/c) \sin \theta_k$ and $k_z = (\omega_\text{in}/c) \cos \theta_k $  denote the components of the linear momentum of the Bessel beam which are perpendicular and parallel, respectively to the beam propagation direction.

The twisted Bessel field [Eqs.~(\ref{A_x})-(\ref{A_z})] can be expressed in the form of Eq.~(\ref{harmonicfield}) and the spatial part $\boldsymbol{A}(\boldsymbol{r})$ can be written in the following manner:
\begin{equation} \label{A_j_r}
    \boldsymbol{A}(\boldsymbol{r}) =  \boldsymbol{g}(\rho,\phi)~e^{i k_z z},
\end{equation}
where $\boldsymbol{g}$ is given by
\begin{equation}
    \begin{split} \label{eq:app_gx}
        g_x =  & \frac{1}{2} A_0 \sqrt{\frac{k_{\perp} } {4\pi} }  \bigg\{ c_{-1} J_{m+1}(k_{\perp} \rho) e^{i(m+1)\phi} \\
        &+ c_{+1} J_{m-1}(k_{\perp} \rho) e^{i(m-1)\phi}
        \bigg\},
    \end{split}
\end{equation}
\begin{equation} \label{eq:app_gy}
    \begin{split}
        g_y =  & \frac{1}{2i} A_0 \sqrt{\frac{k_{\perp} } {4\pi} }  \bigg\{ c_{-1} J_{m+1}(k_{\perp} \rho) e^{i(m+1)\phi } \\
        &- c_{+1} J_{m-1}(k_{\perp} \rho) e^{i(m-1)\phi}
        \bigg\},
    \end{split}
\end{equation}
\begin{equation} \label{eq:app_gz}
        g_z =  \frac{1}{2i} A_0 \sqrt{\frac{k_{\perp} } {2\pi} } c_{0} J_{m}(k_{\perp} \rho) e^{im\phi}.
\end{equation}
Using the expression for the spatial part of the twisted field [Eq.~(\ref{A_j_r})] in Eq.~(\ref{Scattering_amplitude_general}),
\begin{multline} 
\label{Scattering_amplitude_step1}
F_1 =  \sum_{a}^{n_a} e^{-i \bm{k}_s\cdot \bm{R}_a} \\
\times \bra{\psi_{a}} \sum_j^{n_e(a)} e^{-i \bm{k}_s\cdot \bm{ r}_{aj}} \bm \epsilon^* \cdot \bm{g} e^{i \bm{k}_z \cdot (\bm{R}_a + \bm{r}_{aj})} \ket{\psi_{a}},
\end{multline}
where $\boldsymbol{k}_z = k_z \hat{z}$. This can be further simplified as
\begin{equation} \label{Scattering_amplitude_step2}
\begin{split}
    F_1 = & \sum_{a}^{n_a}\! e^{i \boldsymbol{Q}\cdot \boldsymbol{R}_a}  \bra{\psi_{a}} \! \sum_j^{n_e(a)} \!\! e^{i \boldsymbol{Q}\cdot \boldsymbol{ r}_{aj}} \bm \epsilon^* \cdot \boldsymbol{g}  \ket{\psi_{a}},
\end{split}
\end{equation}
where we define $\boldsymbol{Q} = \boldsymbol{k}_z -\boldsymbol{k}_s$ for a twisted Bessel field.
We now proceed by neglecting the variation of the twisted field in the radial ($\rho$) and azimuthal ($\phi$) directions over the electrons within a single atom, but we include the variation of the field across different atoms in a molecule.
\begin{equation} \label{Scattering_amplitude_step3}
    F_1 = \sum_{a}^{n_a}\! e^{i \boldsymbol{Q}\cdot \boldsymbol{R}_a} \bm \epsilon^* \cdot \boldsymbol{g} (\rho_a, \phi_a)\bra{\psi_{a}} \! \sum_j^{n_e(a)} \!\! e^{i \boldsymbol{Q}\cdot \boldsymbol{ r}_{aj}}   \ket{\psi_{a}}.
\end{equation}
For elastic scattering considered here, $k_s \approx \omega_\text{in}/c$. Then the above expression can be reduced further using the atomic form factors $f_a$ to obtain the twisted x-ray scattering amplitude for a single molecule,
\begin{equation}
    F_1 = \sum_{a}^{n_a} f_a(\boldsymbol{Q}) ~ e^{i \boldsymbol{Q}\cdot \boldsymbol{R}_a}  ~ \bm \epsilon^* \cdot \boldsymbol{g}(\rho_a, \phi_a).
\end{equation}
\\

It is worth pointing out that similar expressions to the ones in this work, can be derived using a classical treatment of x-ray scattering of OAM beams from a charge density, as non-resonant time-independent elastic scattering can be described classically. However, a quantum description is used here as it can be simpler, the approximations more physically intuitive, and it can be generalized for other scenarios later.

\section{Convergence behavior}\label{Appendix_convergence}

\begin{figure*}
\resizebox{150mm}{!}{\includegraphics[]{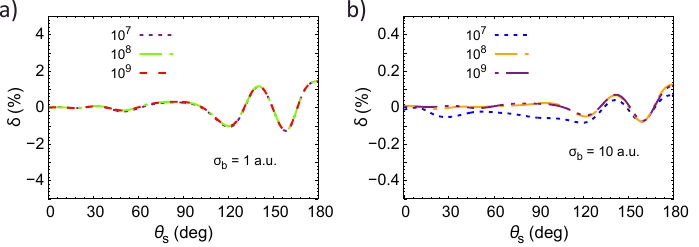}}
\caption{ Convergence behavior of $\delta$ for the single molecule case (Fig.~\ref{fig:single_molecule}c) with increasing number of realizations for different $\sigma_b$;
a) $\sigma_b = 1$ a.u. b) $\sigma_b = 10$ a.u.
}
\label{fig:app_singlmol_convergence}
\end{figure*}

\begin{figure*}
\resizebox{150mm}{!}{\includegraphics[]{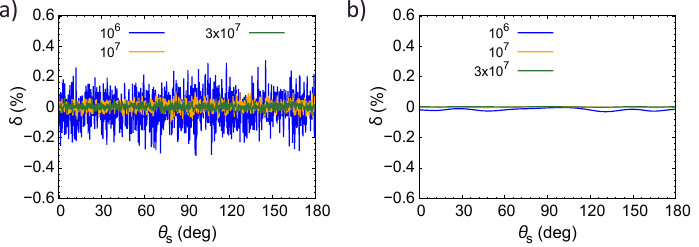}}
\caption{ Convergence behavior of $\delta$ for the 500 molecule ensemble (Fig.~\ref{fig:many_mol}b) with increasing number of ensemble realizations.
a) Coherent sum b) Incoherent sum.
}
\label{fig:app_manymol_convergence}
\end{figure*}

Here, we illustrate the convergence behaviour of the simulations presented in the main text. 
This analysis provides an estimate for the uncertainty in the numerical simulations. 
Figure~\ref{fig:app_singlmol_convergence} shows the single-molecule simulation results from Fig.~\ref{fig:single_molecule}c as a function of the number of realizations. 
Both Fig.~\ref{fig:app_singlmol_convergence}a and \ref{fig:app_singlmol_convergence}b exhibit convergence as the number of realizations approach $10^9$. 
It is also evident that increasing $\sigma_b$ requires a larger number of realizations to adequately sample the phase space. 
The twisted x-ray scattering results from a single molecule reported in the main text are calculated using $10^9$ realizations.

Figure~\ref{fig:app_manymol_convergence} shows the simulation results for the 500 molecule ensemble from Fig.~\ref{fig:many_mol}b as a function of the number of realizations. 
The statistical fluctuations in the coherent sum (Fig.~\ref{fig:app_manymol_convergence}a) decrease with increasing number of ensemble realizations, confirming that the small oscillations about zero arise from finite sampling. 
Figure~\ref{fig:app_manymol_convergence}b reveals that these are absent in the incoherent sum.
This illustrates that the inter-molecular interference terms are the cause of these oscillations which are entirely absent in the incoherent sum calculations.
In addition, the figures show that the incoherent sum converges rapidly, with the dissymmetry factor approaching zero as the number of ensemble realizations increases. 
The molecular ensemble results reported in the main text were calculated using 30 million ensemble realizations.

\section{Dichroic signal in the paraxial approximation} \label{Appendix_paraxial_approx}
In this section, we derive the dissymmetry factor $\delta$ analytically for some special cases with additional approximations to offer further insights into the simulations presented in the main text. 
In the paraxial approximation, for $\Lambda = 1$, the twisted Bessel field $\bm{g}$ is
\begin{equation} \label{paraxial_g_x}
    \begin{split}
        g_x \approx  & \frac{1}{2} A_0 \sqrt{\frac{k_{\perp} } {4\pi} } J_{m-1}(k_{\perp} \rho) e^{i(m-1)\phi},
    \end{split}
\end{equation}
\begin{equation} \label{paraxial_g_y}
    \begin{split}
        g_y \approx  & \frac{i}{2} A_0 \sqrt{\frac{k_{\perp} } {4\pi} } J_{m-1}(k_{\perp} \rho) e^{i(m-1)\phi},
    \end{split}
\end{equation}
\begin{equation} \label{paraxial_g_z}
        g_z \approx 0.
\end{equation}

For forward scattering ($\theta_s = 0$), two allowed choices for the outgoing photon polarization are $\bm \epsilon_1 = (0,1,0)$ and $\bm \epsilon_2 = (-1, 0 , 0)$.                                                              
Using Eq.~(\ref{Scattering_amplitude_molecule}), the scattering probability amplitude for a single molecule is given by    
\begin{align} 
F(\bm \epsilon_1) &= \sum_{a}^{n_a} f_a(0) ~  \boldsymbol{g}_y(\rho_a, \phi_a), \label{forward_amp_e1}\\
F(\bm \epsilon_2) &= -\sum_{a}^{n_a} f_a(0) ~  \boldsymbol{g}_x(\rho_a, \phi_a).\label{forward_amp_e2}
\end{align}

For small cone angles, the variation of intensity over a given molecule is negligible, therefore the approximate intensity computed at the center of mass of the molecule is used. 
Substituting the expressions for $\bm g$ from Eqs.~(\ref{paraxial_g_x})-(\ref{paraxial_g_z}) in Eq.~(\ref{forward_amp_e1}),
\begin{equation} \label{forward_amp_e1_step}
\begin{split}
    F(\bm \epsilon_1) = & \frac{i}{2} A_0 \sqrt{\frac{k_{\perp} } {4\pi} } J_{m-1}(k_{\perp} \rho) \sum_{a}^{n_a} f_a(0) e^{i(m-1)\phi_a}.
\end{split}
\end{equation}
As the form factors are real for non-resonant elastic scattering, the scattering yield is given by
\begin{equation}
    \begin{split}
        S(\bm \epsilon_1) = \abs{F(\bm \epsilon_1)}^2 = & \frac{1}{4} A_0^2 \frac{k_{\perp} } {4\pi}  J_{m-1}^2(k_{\perp} \rho) \bigg[ \sum_{a} \abs{ f_a(0)}^2\\
        & + \sum_{a \neq b} f_a(0) f_b(0) e^{i(m-1)(\phi_b - \phi_a )} \bigg].
    \end{split}
\end{equation}
Since in the paraxial approximation, the difference in the yield from two enantiomers is approximately equal to the difference in the scattering yield between two complementary TAM values $(m,2-m)$ for the same enantiomer we get,
\begin{widetext}

\begin{multline} \label{eq:paraxial_forward_pol1}
S_m(\bm \epsilon_1)  - S_{2-m}(\bm \epsilon_1)
 =  \frac{1}{4} A_0^2 \frac{k_{\perp} } {4\pi}  J_{m-1}^2(k_{\perp} \rho) 
\Bigg[ \sum_{a \neq b}  f_a(0) f_b(0) \big[ e^{i(m-1)(\phi_b - \phi_a )} - e^{i(1-m)(\phi_b - \phi_a )} \big] \Bigg] \\
 =  \frac{1}{4} A_0^2 \frac{k_{\perp} } {4\pi}  J_{m-1}^2(k_{\perp} \rho) \Bigg[ \sum_{a > b}  f_a(0) f_b(0) (2i) \bigg\{ \sin{\big[(m-1)(\phi_b - \phi_a )\big]} + \sin{\big[(m-1)(\phi_a - \phi_b )\big]}  \bigg\} \Bigg] \ = 0
\end{multline}
\end{widetext}
Here, we have used $J_{m-1}^2(k_{\perp} \rho) = J_{1-m}^2(k_{\perp} \rho) $. Similarly one can also prove
\begin{equation}
    S_m(\bm \epsilon_2)  - S_{2-m}(\bm \epsilon_2) = 0.
\end{equation}

\section{Intensity of twisted Bessel field} \label{Appendix_Intensity_twisted}
The intensity of the twisted beam has been previously derived by Surzhykov et al.~\cite{surzhykov2016probing} and shown to be the component of the Poynting vector along the propagation direction.
The instantaneous Poynting vector $\mathcal{S}$ is defined as
\begin{equation}
    \mathcal{S}(t)  = \frac{1}{\mu_0} \bm E (t) \times \bm B(t).
\end{equation}
where the electric field $\bm E = -\partial \bm A_\text{in} / \partial t$ and the magnetic field $\bm B = \curl{\bm A_\text{in}}$.
Using Eq.~(\ref{harmonicfield}),
\begin{widetext}

\begin{multline}
\mathcal{S}(t) =  
\bigg(\frac{c^2}{4\pi}\bigg) i\omega_\text{in} 
\bigg( \bm A \times (\curl{\bm A}) e^{-2i\omega_\text{in}t} 
+ \bm A \times (\curl{\bm A^*}) 
- \bm A^* \times (\curl{\bm A}) 
- \bm A^* \times (\curl{\bm A^*}) e^{2i\omega_\text{in}t} \bigg).   
\end{multline}

The time-averaged Poynting vector is
\begin{equation} \label{eq:timeavg_poynting}
     \langle \mathcal{S} \rangle = \bigg(\frac{c^2}{4\pi}\bigg) i\omega_\text{in} \bigg( \bm A \times (\curl{\bm A^*}) - \bm A^* \times (\curl{\bm A})  \bigg ).
\end{equation}
The intensity is given by $I = \abs{ \langle \mathcal{S} \rangle _z}$. 
Using the expression for the twisted Bessel field Eqs.~(\ref{A_j_r})-(\ref{eq:app_gz}) , the $z$-component of the right-hand side of Eq.~(\ref{eq:timeavg_poynting}) is
\begin{equation} \label{eq:rhs_S_st1}
\begin{split}
     \bigg( \bm A & \times (\curl{\bm A^*}) 
     - \bm A^* \times (\curl{\bm A})  \bigg )_z \\
      & =  \Bigg(\frac{-i}{16\pi}\Bigg) A_0^2 \Bigg[ k_{\perp} \bigg( 4c^2_{+1} k_z - \sqrt 2 ~ c_0 ( c_{-1} - 2 c_{+1} ) k_{\perp} \bigg) J^2_{m-1} + \frac{4\sqrt{2}~ c_0 ~ c_{-1} k_{\perp} m J_{m-1}J_m}{\rho} \\
      & ~~~ - \frac{4\sqrt 2 ~ c_0 ~ c_{-1} m^2 J^2_m}{\rho^2} 
       + c_{-1}k_{\perp} \bigg(4c_{-1}k_z - \sqrt 2 ~ c_0 ~ k_{\perp} \bigg) J^2_{m+1}  \Bigg].
\end{split}
\end{equation}
Using the recursive Bessel relation $\frac{2m}{x} J_m(x) = J_{m-1}(x) + J_{m+1}(x)$, 
\begin{equation} \label{eq:rhs_S_st2}
\begin{split}
     \bigg( \bm A & \times (\curl{\bm A^*}) 
     - \bm A^* \times (\curl{\bm A})  \bigg )_z \\
     & = \Bigg( \frac{-i}{8\pi} \Bigg) A^2_0 k_{\perp} \Bigg[ \bigg(  2c^2_{+1} k_z + \sqrt 2 c_0 c_{+1} k_{\perp} \bigg)  J^2_{m-1} + \bigg( 2c^2_{-1} k_z - \sqrt 2 c_0 c_{-1} k_{\perp} \bigg) J^2_{m+1} \Bigg].
\end{split}
\end{equation}
The quantities on the right-hand side inside the parenthesis can be further simplified using the expressions for $k_{\perp}, k_z$, Eq.~(\ref{eq:defn_c_coeff}) and using $\Lambda^2 = 1$,
\begin{equation} \label{eq:rhs_S_st3}
2c^2_{+1} k_z + \sqrt 2 c_0 c_{+1} k_{\perp}  = 2 \frac{\omega_\text{in}}{c} \Lambda ~ c^2_{+1},
\end{equation}
\begin{equation} \label{eq:rhs_S_st4}
2c^2_{-1} k_z - \sqrt 2 c_0 c_{-1} k_{\perp} = -2 \frac{\omega_\text{in}}{c} \Lambda ~ c^2_{-1}.
\end{equation}

Combining the results from Eqs.~(\ref{eq:rhs_S_st2}) -(\ref{eq:rhs_S_st4}), the intensity is given by
\begin{equation} \label{eq:timeavg_poynting_final}
     I = \Bigg(\frac{c}{4\pi}\Bigg) \Bigg[ \frac{A^2_0 k_{\perp} \omega^2_\text{in} }{4\pi} \Big| c^2_{+1} J^2_{m-1} (k_\perp \rho) - c^2_{-1} J^2_{m+1}(k_\perp \rho) \Big| \Bigg].
\end{equation}
\noindent where atomic units have been used throughout ($\mu_0 = 4\pi/c^2$).

\end{widetext}

\bibliographystyle{unsrt}
\bibliography{biblio}

\end{document}